\documentclass[sigconf,9.99pt]{acmart}
\usepackage{graphicx} % Required for inserting images
\usepackage{subfig}

\usepackage{hyperref}
\makeatletter
\def\UrlAlphabet{%
      \do\a\do\b\do\c\do\d\do\e\do\f\do\g\do\h\do\i\do\j%
      \do\k\do\l\do\m\do\n\do\o\do\p\do\q\do\r\do\s\do\t%
      \do\u\do\v\do\w\do\x\do\y\do\z\do\A\do\B\do\C\do\D%
      \do\E\do\F\do\G\do\H\do\I\do\J\do\K\do\L\do\M\do\N%
      \do\O\do\P\do\Q\do\R\do\S\do\T\do\U\do\V\do\W\do\X%
      \do\Y\do\Z}
\def\UrlDigits{\do\1\do\2\do\3\do\4\do\5\do\6\do\7\do\8\do\9\do\0}
\g@addto@macro{\UrlBreaks}{\UrlOrds}
\g@addto@macro{\UrlBreaks}{\UrlAlphabet}
\g@addto@macro{\UrlBreaks}{\UrlDigits}
\makeatother

\AtBeginDocument{ \providecommand\BibTeX{{ \normalfont B\kern-0.5em{\scshape i\kern-0.25em b}\kern-0.8em\TeX}}}

\setcopyright{acmlicensed}
\copyrightyear{2025}
\acmYear{2025}
\acmDOI{10.1145/xxxxxx.yyyyyyy}

\acmConference[MobiSys '25]{The 23nd ACM International Conference on Mobile Systems, Applications, and Services}{June 23--27, 2025}{Anaheim, CA, USA}

\begin{document}

\title{Demo: Cellular-X: An LLM-empowered Cellular Agent for Efficient Base Station Operations}

\author{Liujianfu Wang}
\affiliation{ \institution{The Chinese University of Hong Kong} \city{Hong Kong SAR}\country{China}}
%\affiliation{ \institution{CUHK} \city{Hong Kong SAR}\country{China}}
\email{1155157110@link.cuhk.edu.hk}
\authornote{Authors contributed equally to this research.}

\author{Xinyi Long}
%\affiliation{ \institution{CUHK} \city{Hong Kong SAR}\country{China}}
\affiliation{ \institution{The Chinese University of Hong Kong} \city{Hong Kong SAR}\country{China}}
\email{1155222493@link.cuhk.edu.hk}
\authornotemark[1]

\author{Yuyang Du}
%\affiliation{ \institution{CUHK} \city{Hong Kong SAR}\country{China}}
\affiliation{ \institution{The Chinese University of Hong Kong} \city{Hong Kong SAR}\country{China}}
\email{dy020@ie.cuhk.edu.hk}
\authornotemark[1]

\author{Xiaoyan Liu}
%\affiliation{ \institution{CUHK} \city{Hong Kong SAR}\country{China}}
\affiliation{ \institution{The Chinese University of Hong Kong} \city{Hong Kong SAR}\country{China}}
\email{1155223027@link.cuhk.edu.hk}

\author{Kexin Chen}
%\affiliation{ \institution{CUHK} \city{Hong Kong SAR}\country{China}}
\affiliation{ \institution{The Chinese University of Hong Kong} \city{Hong Kong SAR}\country{China}}
\email{1155184119@link.cuhk.edu.hk}

\author{Soung Chang Liew}
%\affiliation{ \institution{CUHK} \city{Hong Kong SAR}\country{China}}
\affiliation{ \institution{The Chinese University of Hong Kong} \city{Hong Kong SAR}\country{China}}
\email{soung@ie.cuhk.edu.hk}

\begin{abstract}
This paper introduces \textit{Cellular-X}, an LLM-powered agent designed to automate cellular base station (BS) maintenance. Leveraging multimodal LLM and retrieval-augmented generation (RAG) techniques, \textit{Cellular-X} significantly enhances field engineer efficiency by quickly interpreting user intents, retrieving relevant technical information, and configuring a BS through iterative self-correction. Key features of the demo include automatic customized BS setup, document-based query answering, and voice-controlled configuration reporting and revision. We implemented \textit{Cellular-X} on a USRP X310 testbed for demonstration. Demo videos and implementation details are available at \textcolor{blue}{\url{https://github.com/SeaBreezing/Cellular-X}}.
\end{abstract}

\begin{CCSXML}
<ccs2012>
   <concept>
       <concept_id>10003033.10003058</concept_id>
       <concept_desc>Networks~Network components</concept_desc>
       <concept_significance>300</concept_significance>
       </concept>
   <concept>
       <concept_id>10003120.10003138</concept_id>
       <concept_desc>Human-centered computing~Ubiquitous and mobile computing</concept_desc>
       <concept_significance>500</concept_significance>
       </concept>
 </ccs2012>
\end{CCSXML}

\ccsdesc[500]{Human-centered computing~Ubiquitous and mobile computing}
\ccsdesc[300]{Networks~Network components}

\keywords{Cellular network, base station management, LLM, RAG, USRP}
  
\maketitle

\section{Introduction} \label{sec1}
With recent trends toward denser and smaller cells driven by the millimeter-wave technology, a larger-scale deployment of cellular base stations (BSs) is essential to the evolution of modern mobile systems. Managing the growing number of increasingly complex BSs is a daunting task. Manual BS maintenance is both labor-intensive and error-prone. Engineers often spend exorbitant amounts of time consulting technical documents and writing custom scripts to configure and operate a BS, which we believe could be effectively automated with AIs.

This paper introduces \textit{Cellular-X}, an LLM-powered cellular network agent that significantly accelerates standard document retrieval and BS maintenance. Leveraging a multimodal LLM and the retrieval-augmented generation (RAG) technique, the agent interprets human intents and efficiently retrieve relevant information from voluminous and complex technical documents. Further, the agent automates the generation of configuration files in an iterative manner, continuously analyzing error logs for self-correction and engaging with users to refine instructions as needed.

\textbf{Related Works}: Many recent studies have explored the application of LLMs in wireless networking and mobile systems. However, most of them focus on management and optimization tasks for backbone networks, with relatively few addressing BS-related challenges. For example, \cite{kouachi2024leveraging} examined LLM-based traffic analysis at BSs, a 3GPP standard analysis framework using LLMs was proposed in \cite{nikbakht2024tspec}, and the BS siting problem was explored in \cite{wang2025large}. Notably, none of these works have demonstrated real-world systems, and to our best knowledge, no prior study has investigated the use of LLMs for autonomous BS maintenance.

\section{System Overview} \label{sec2}
\textit{Cellular-X} assists field engineers with three major functionalities: 1) automatic BS configuration with customized setup, 2) answering the engineer's questions according to technical or standard-related documents, and 3) voice-controlled configuration report and modification. This section provides a high-level system overview, and we refer readers to our GitHub repository for implementation details. In general, \textit{Cellular-X} consists of the following subsystems:

%Before comprehensively testing these functions and presenting associated demonstrations in Section \ref{sec3}, this section provides a system design overview; we refer readers to our GitHub repository for implementation details. From a high-level perspective, \textit{Cellular-X} consists of the following subsystems.

%From a high-level perspective, \textit{Cellular-X} consists of the following subsystems

The \textbf{Configuration Subsystem} aims to generate proper BS configurations according to users' customized requirements. There are two major functional units to consider when setting a BS - Evolved Packet Core (EPC) and Evolved NodeB (ENB). To reduce the task complexity, we divide the BS configuration process into two subtasks, with each one handling a functional unit exclusively. Further, the processing of each subtask comprises two phases: the initialization phase and the self-correction phase. In the initialization phase, user requirements are conveyed to the agent via the user's input prompt. Historical configuration files of the unit, along with their resulting system logs, are added to the prompt for the configuration file's generation. To accelerate the initialization process and avoid past mistakes, the subsystem builds the initial configuration file based on the most similar configuration from previous records. The initialization phase is followed by an iterative self-correction process, in which the latest system log of a BS unit is analyzed during each iteration to ensure the configuration is progressively refined toward the desired cellular network setup.

\begin{figure}
    \centering
    \includegraphics[width=0.40\textwidth]{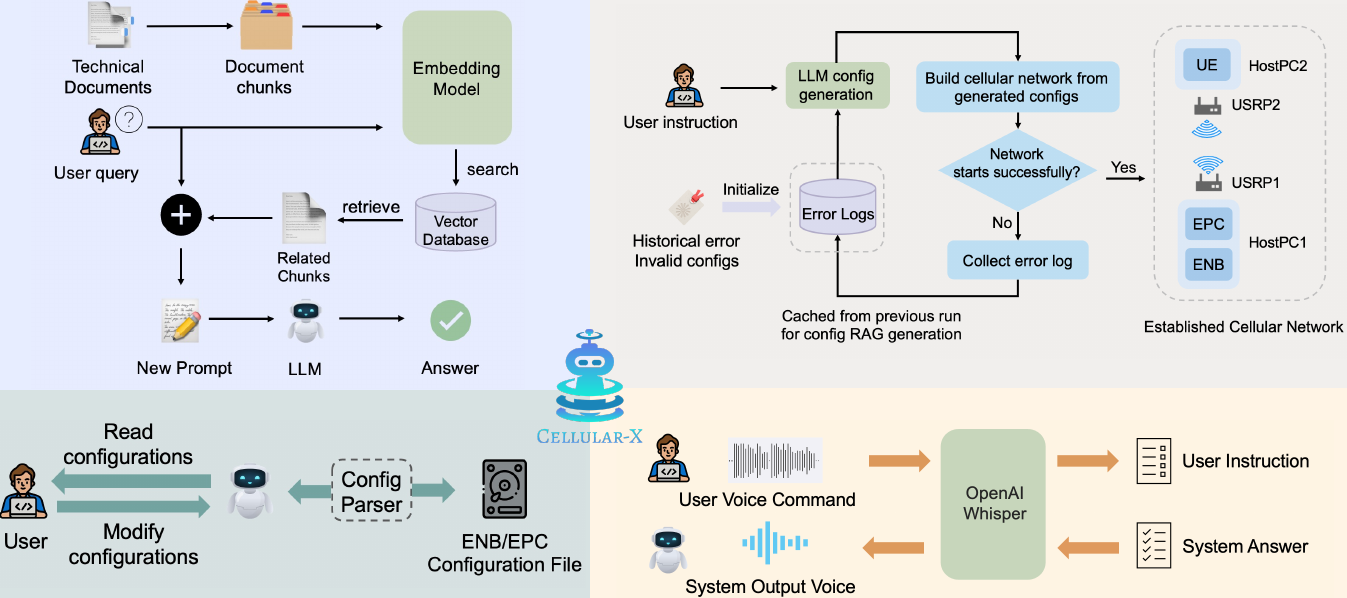}
    \caption{A systematic overview of Cellular-X, in which four subsystems work cooperatively for the designed functions.}\vspace{-1.5em}
    \label{fig:subsystems}
\end{figure}

The \textbf{RAG Subsystem} is designed to save engineers' document search time when maintaining a BS. When a user poses a question to the agent, such as the allowable range of a parameter in the 3GPP standard, the system responds with a vocal answer. To obtain the most accurate answer, the RAG subsystem retrieves relevant information from technical specifications, such as standard documents and technical handbooks, and combines the retrieved material with the user's query to form a new prompt for the agent's subsequent processing. The retrieval begins by processing the user's input with an embedding network. This is followed by matching between the embedding of the user's input and the embeddings of pre-chunked technical documents. The matching process follows a ``Top-k" policy, in which the $k$ most similar document chunks (measured in terms of cosine similarity) are retrieved as reference information.

The \textbf{File Read/Write Subsystem} is granted read/write access to BS configuration files so that it can provide the cellular agent with the ability to query/modify BS settings based on user instructions.

The \textbf{Human Interaction Subsystem} facilitates an engineer's voice-based system control with the integration of a general-purpose multimodal LLM, which converts user voice commands into text prompts and delivers audio responses as system output.

\section{Experiment and Demonstration} \label{sec3}
\begin{figure}
    \centering
    \includegraphics[width=0.394\textwidth]{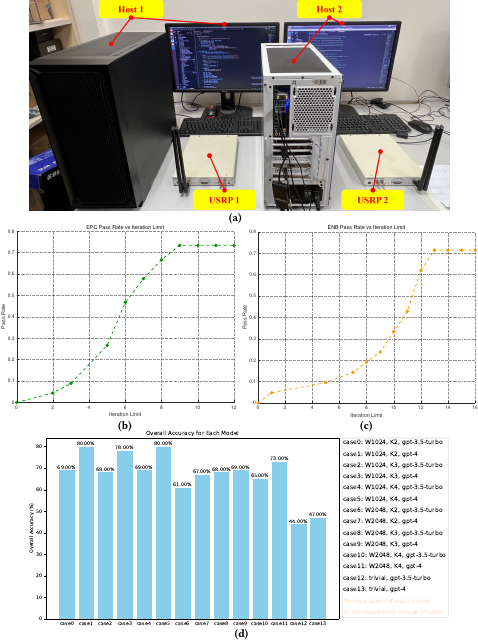}
    \caption{Experimental details and testing results}\vspace{-1.5em}
    \label{fig:experiments}
\end{figure}

For a comprehensive demonstration, we deploy \textit{Cellular-X} in the software-defined radio (SDR) testbed presented in Fig. \ref{fig:experiments}(a). For the implementation of \textit{Cellular-X}, we use Claude-3.5 Sonnet as the underlying model for both the configuration and file read/write subsystems, and GPT-4 for the RAG subsystem. The human interaction subsystem is powered by the OpenAI Whisper model. We build a practical BS and an associated user equipment (UE) using two USRP X310, with the srsRAN LTE project running on each device. Each USRP's host PC is powered by an AMD Ryzen 1950X processor and the Ubuntu 18.04 operating system.

We present three representative demonstration videos, available on our \textit{Cellular-X}'s GitHub project and YouTube channel.\footnote{\textcolor{blue}{\url{https://youtube.com/playlist?list=PLi7wIohZ9VLjfbtShawzEk49BKUE11QiU&}}}

In \textbf{Demo \#1}, the user issues a voice command instructing the agent to set up the desired LTE BS and connect the UE to the cellular network. Fig. \ref{fig:experiments}(b) and Fig. \ref{fig:experiments}(c) present the agent's performance in configuring the EPC and ENB units under varying self-correction iteration limits. The figures show that while increasing allowed iterations improves the success probability, this improvement has an upper limit, probably due to potential errors in the initial configuration, i.e., if the initial configuration contains severe error(s), further self-correction may not always helpful in terms of the BS's successful establishment. However, this problem can be partially mitigated as the agent accumulates more experience in BS maintenance, gradually building a richer database of previously successful configurations with fewer errors. \textbf{Demo \#2} demonstrates the agent's capability to assist users by promptly and precisely answering technical questions. In this demo, we utilize TSpec-LLM \cite{nikbakht2024tspec}, a 3GPP documentation database, as the RAG library, and OpenAI's \textit{text-embedding-ada-002} as the embedding network for similarity calculations. Fig. \ref{fig:experiments}(d) shows significant accuracy improvements compared to trivial methods without RAG. Finally, \textbf{Demo \#3} showcases the agent's capability in reporting the latest configuration parameters and revising them based on the user's voice instructions.

\bibliographystyle{ACM-Reference-Format}
\bibliography{references}
\end{document}